\documentclass[twocolumn,pra,showpacs,superscriptaddress]{revtex4}
\usepackage{graphicx}
\usepackage{amsmath}
\begin{document}

\title{Relaxation oscillations in long-pulsed random lasers}
\date{\today}
\author{Karen L. van der Molen}
\email{k.l.vandermolen@utwente.nl} \affiliation{Complex Photonic
Systems, MESA$^+$ Institute for Nanotechnology and Department of
Science and Technology\\ University of Twente, PO Box 217, 7500 AE
Enschede, The Netherlands.}
\author{Allard P. Mosk}
\affiliation{Complex Photonic Systems, MESA$^+$ Institute for
Nanotechnology and Department of Science and Technology\\ University
of Twente, PO Box 217, 7500 AE Enschede, The Netherlands.}
\author{Ad Lagendijk}
\affiliation{FOM Institute for Atomic and Molecular Physics (AMOLF),
Kruislaan 407, 1098 SJ Amsterdam, The Netherlands}

\begin{abstract}
We have measured the evolution of the light intensity of a random
laser during a nanosecond pump pulse. Relaxation oscillations in a
titania random laser were observed in the time trace of the total
emitted intensity. We compare our experimental results with a simple
model, based on the four-level rate equations for a single mode
laser.
\end{abstract}


\maketitle

\section{Introduction}

Relaxation oscillations of conventional lasers are a well-understood
phenomenon.\cite{Siegman1986} They are especially important for
continuous wave and long-pulsed lasers. In a random laser, a medium
in which gain is combined with multiple scattering of light,
relaxation oscillations can also occur, as was predicted by Letokhov
in 1968.\cite{Letokhov1968} A preeminent experimental demonstration
of a random laser was published by Lawandy \textit{et al.} in 1994
\cite{Lawandy1994}, followed by many others
\cite{Sharma2006,Rose2005,Cao1998,Mujumdar2004}. Only recently
Soukoulis and coauthors have presented measurements of relaxation
oscillations in single modes of a picosecond pumped random laser
system.\cite{Soukoulis2002} Different numerical calculations on
random lasers also show this oscillatory
behavior.\cite{Wiersma1996,Berger1997,Noginov2004} To our knowledge,
no measurements have been performed on relaxation oscillations in
random lasers in the interesting regime of long pulses.

In this report we present measurements of the time evolution of a
nanosecond pumped random laser system. We compare our experimental
observations with a simple model, based on the four-level rate
equations for a single-mode laser.

\section{A simple model \label{ch_relosc_sec,th_sim_model}}
The evolution of the excited molecules $N_1$ and the number of
photons $q$ in a laser are described by the well-known four-level
rate equations \cite{Siegman1986}
\begin{subequations}\label{ch_rlintro_eqn,rate}
\begin{align}
&\frac{dN_1(t)}{dt}  =  P(t) - \frac{\beta
q(t)N_1(t)}{\tau}-\frac{N_1(t)}{\tau}, \label{ch_rlintro_eqn,rateN1}
\\
&\frac{dq(t)}{dt}  =  - \frac{q(t)}{\tau_c} +\frac{\beta
N_1(t)}{\tau}\left[ q(t) + 1 \right], \label{ch_rlintro_eqn,rateq}
\end{align}
\end{subequations}
where $P$ is the pumping fluence that is absorbed by the molecules
inside the cavity, $\tau_c$ the cavity decay time, $\tau$ the
spontaneous-emission life time, and $\beta$ the beta factor of the
laser. The beta factor is defined as the amount of spontaneous
emission that contributes to the lasing mode, and can also be
determined for a random laser~\cite{Soest2002}.

From these rate equations Woerdman and coauthors have derived an
equation for the frequency of the relaxation oscillations explicitly
including the spontaneous emission:\cite{Woerdman2001a}
\begin{eqnarray}
\omega_{\rm{res}} = \sqrt{\left (\frac{M - 1}{\tau_c \tau} \right
)-\frac{1}{4}\left [ \frac{M}{\tau} - \frac{\beta}{\tau_c (M - 1)}
\right ]^2}, \label{ch_relosc_eqn,fitoscillations}
\end{eqnarray}
where $\omega_{\rm{res}}$ is the relaxation oscillations frequency,
and $M$ the scaled pump fluence, defined as the ratio of the
absorbed fluence $P$ and the threshold fluence $P_{th}$.

We apply this model to the multi-mode pulsed random laser by simply
changing $\beta$ to $\beta_{\rm{mm}}$ and $\tau_c$ to
$\tau_{\rm{c,mm}}$, i.e. we use a mean cavity decay time and a mean
beta factor to describe our multi-mode random laser. This
simplification accurately describes the threshold behavior of random
lasers~\cite{Molen2006a}.

\section{Experimental apparatus \label{ch_relosc_sec,exp_app}}
The random laser used in our experiments consists of a suspension of
TiO$_2$ particles (mean diameter of 180~nm, volume fraction of 10\%)
in a solution of Sulforhodamine B in methanol (1~mmol/liter; pump
absorption length, 104~$\mu$m; minimal gain length, 83~$\mu$m
\cite{Soest2001}). The suspension is contained in a fused silica
capillary tube, with internal dimensions $100 \times 2 \times
2$~mm$^3$. To measure the mean free path of light in this sample, we
performed an enhanced-backscatter cone experiment \cite{Albada1985}
and an escape function experiment \cite{Gomez2003}. We found a
transport mean free path of $0.46\pm0.1$ $\mu$m at 633~nm (effective
refractive index, 1.48$\pm$0.04).

The samples were excited by a pump pulse at 532~nm, provided by an
optical parametric oscillator (OPO) pumped by a Q-switched Nd:YAG
laser (Coherent Infinity 40-100/XPO). The pump pulse had a duration
of 3~ns and a repetition rate of 50~Hz. The pump light was focused
with a microscope objective (water-immersed, numerical aperture
NA,~1.2) onto the sample (focus area, $12 \pm 6$~$\mu$m$^2$),
reaching an intensity in the order of 1~mJ/mm$^2$. The central
wavelength of the emitted light is 594~nm \cite{Molen2006a}, and the
narrowing factor (defined as the spectral width of the emitted light
far above threshold divided by the spectral width far below
threshold) is 8. The light emitted by the random laser was collected
by the same microscope objective. The emitted light was detected by
a 25~GHz photodiode (New Focus 1404), read out by an oscilloscope
(Tektronix TDS 7404, analog bandwidth, 4~GHz). The resulting time
resolution was 100~ps. To obtain a good signal-to-noise ratio, the
data shown is an average of 100 oversampled time traces. The pump
light was filtered out of the detection path by use of a colored
glass filter with an optical density of more than 4 at the
wavelength of the pump laser.

\section{Measured relaxation oscillations \label{ch_relosc_sec,exp_results}}
The normalized time trace of the pump pulse and the normalized time
trace of the total emitted light from the random laser far above
threshold are shown in Fig.~\ref{ch_relosc_fig,pumppulse}. Overall,
the duration of the pump laser pulse is longer than the duration of
the pulse of light the random laser emits. We see in the pulse
emitted by the random laser first a fast decay, followed by a slower
exponential decay. The fast decay is due to the stimulated emission
in the random laser. In the second part of the decay the population
inversion is no longer present, and the spontaneous emission causes
a slower decay of intensity. These observations are in agreement
with other random laser experiments.\cite{Soukoulis2002} In
Fig.~\ref{ch_relosc_fig,pumppulse} relaxation oscillations in the
emitted light are clearly visible near the peak intensity.

\begin{figure}
\begin{center}
\includegraphics[width=3.4in]{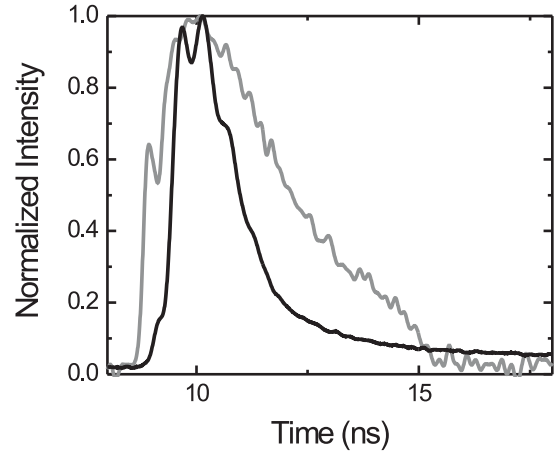}
\end{center}
\caption{\label{ch_relosc_fig,pumppulse}Measured time traces of the
pump pulse (gray) and emission output above threshold (black, input
fluence = 0.47~mJ/mm$^2$) of a titania random laser. The pump pulse
duration is much longer than the duration of the emitted light of
the random laser above threshold. Relaxation oscillations in the
emitted light are clearly visible near the peak intensity. The decay
time of the emitted light is first dominated by stimulated emission.
In the second part of the decay-curve, the spontaneous emission is
dominating.}
\end{figure}

We measured the time evolution for different input fluences. In
Fig.~\ref{ch_relosc_fig,temporalevolution} the normalized intensity
is plotted versus time for four different pump fluences \footnote{A
slight time shift is possible, since the triggering of the
oscilloscope was found to depend on the pump laser intensity.}. The
time traces are shifted vertically with respect to each other for
clarity. The time trace at a pump fluence of 0.06~mJ/mm$^2$ is below
threshold, while the time traces with higher pump fluences are above
threshold. We observe that relaxation oscillations occur above
threshold and become more pronounced when the pump fluence
increases.

\begin{figure}
\begin{center}
\includegraphics[width=3.4in]{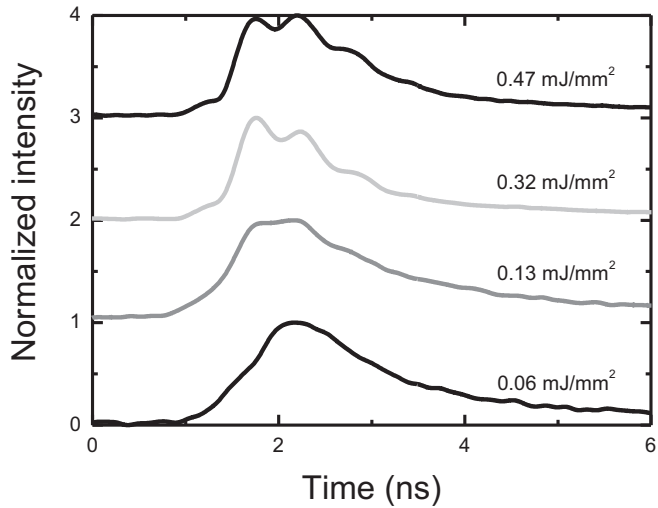}
\end{center}
\caption{\label{ch_relosc_fig,temporalevolution}Measured time traces
of the emission intensities of a titania random laser for four
different pump fluences. The traces are vertically shifted with
respect to each other for clarity. Relaxation oscillations become
more pronounced at higher pump fluences.}
\end{figure}

The frequency of the relaxation oscillations are computed from the
time traces. We determine the times at which the intensity is at a
local maximum. The difference of two consecutive local maxima
$\Delta t$ is the period, and the frequency of the relaxation
oscillation $\nu_{\rm{rel}}$ is given by $ 1 / (\Delta t)$.

\section{Comparison of the measurements with the model \label{ch_relosc_sec,comp_model_exp}}
\begin{figure}
\begin{center}
\includegraphics[width=3.4in]{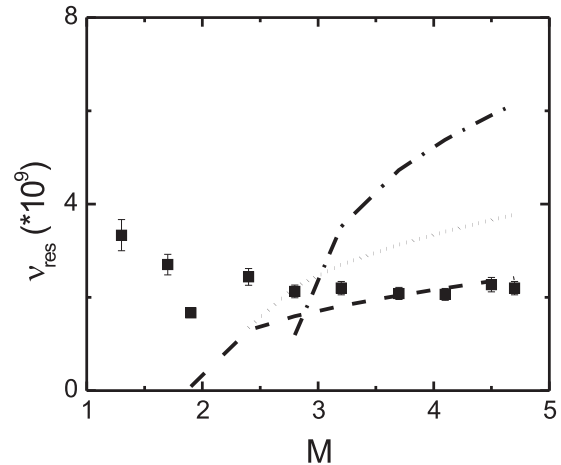}
\end{center}
\caption{\label{ch_relosc_fig,fitmeasurement}Measured relaxation
oscillations frequency as a function of the normalized pump fluence
(squares). The simple model
[Eq.~(\ref{ch_relosc_eqn,fitoscillations})] is plotted for different
cavity decay times: 0.7 (dash-dotted line), 2 ps (dotted line), and
5 ps (dashed line). The fit of the model for a cavity decay time of
5~ps fits reasonably for high pump fluences, but corresponds to a
surprisingly large value of $\tau$.}
\end{figure}
We have inferred the relaxation oscillation frequency for different
pump fluences from our measurements. In
Fig.~\ref{ch_relosc_fig,fitmeasurement} the measured relaxation
oscillations frequencies are plotted versus the scaled pump fluence
$M$. The relaxation-oscillation frequency significantly decreases
when the scaled pump fluence increases from 1~to~2. A further
increase of the scaled pump fluence does not significantly change
the frequency of the relaxation oscillations. The result of
Eq.~(\ref{ch_relosc_eqn,fitoscillations}) is depicted for different
cavity decay times. This cavity decay time is the only parameter
that could not be directly determined by our experiment. The trend
of the model for a fixed cavity decay time is that, in contrast to
our measurements, the relaxation-oscillations frequency increases
for increasing pump fluence. Only for large ($>$~3) normalized pump
fluence the fit of the model for a cavity decay time of 5~ps fits
reasonably.

The difference in the observed trend of the relaxation-oscillation
frequency between the simple model and our experiments, is probably
due to the size of the gain volume: The gain inside the gain volume
of the random laser saturates at threshold \cite{Molen2006a}. A
further increase of the pump fluence will lead immediately to a
larger gain volume, and a corresponding longer cavity decay time.
Apparently, this behavior saturates above a scaled pump fluence of
3.

We have used the distribution of the phase-delay time to determine
the mean cavity decay time $\overline{\tau_c}$, given by
\cite{Molen2006a}
\begin{eqnarray}
\overline{\tau_c} = \frac{1}{8} \frac{L^2}{D},
\label{ch_relosc_eqn,tauc}
\end{eqnarray}
with $L$ the length of the gain volume and, for non-resonant
scattering, the diffusion constant $D = c_0 \ell / (3 n')$, where
$c_0$ is the speed of light in vacuum, $\ell$ the transport mean
free path and $n'$ is the real part of the effective refractive
index of the medium. For our titania random laser we find a
$\overline{\tau_c}$ in the order of 10$^{-13}$~s, in contrast to
10$^{-12}$~s that is suggested by the agreement in
Fig.~\ref{ch_relosc_fig,fitmeasurement} at $M=3$. The difference
between the two cavity decay times is a factor 10. This deviation
could originate in part from the difference between the mean cavity
decay time $\overline{\tau_c}$ of all modes and the mean cavity
decay time of the lasing modes: Our experiment suggests that random
lasing preferentially takes place in modes with a much longer than
average cavity decay time.

\section{Conclusions \label{ch_relosc_sec,conclusions}}
We have seen relaxation oscillations in our random laser, while
looking at the time evolution of the total emitted light for
different realizations of the sample. Multiple modes contributed to
these time traces, and we averaged the time traces over several
realizations of disorder of our random laser sample. The resulting
time trace still showed relaxation oscillations: a weighted average
of the oscillations of all the underlying modes.

The measured relaxation oscillations were compared with a simple
model, based on a single-mode continuous-wave laser system. The
observed trend of our experiments differs from the expected trend,
due to the increase of the gain volume and the corresponding cavity
decay time for an increase of the scaled pump fluence. The cavity
decay time determined with the fit from the simple model is a factor
10 higher than the mean cavity decay time of our sample. Our
experiment suggests that the modes contributing to random laser
emission have a cavity decay time much longer than the average
cavity decay time.

\section*{Acknowledgements}
This work is part of the research program of the 'Stichting voor
Fundamenteel Onderzoek der Materie' (FOM), which is financially
supported by the 'Nederlandse Organisatie voor Wetenschappelijk
Onderzoek' (NWO). We thank Boris Bret for the work that lead to
the estimation of the transport mean free path.

\end{document}